\documentclass[onecolumn,letterpaper,amsmath,amssymb,floatfix,aps,superscriptaddress]{revtex4}
\usepackage{subfigure}
\usepackage{color}

\usepackage{graphicx}
\usepackage{bm}  
\usepackage{epsfig}
\usepackage{mathtools}
\usepackage{verbatim} 

\begin{document}

\title{Correlation functions of main-chain polymer nematics constrained by tensorial and vectorial conservation laws}


\author{Daniel Sven\v sek}
\affiliation{Department of Physics, Faculty of Mathematics and Physics, University of Ljubljana, Jadranska 19, SI-1000 Ljubljana, Slovenia}

\author{Rudolf Podgornik}
\affiliation{Department of Physics, Faculty of Mathematics and Physics, University of Ljubljana, Jadranska 19, SI-1000 Ljubljana, Slovenia}
\affiliation{Department of Theoretical Physics, J. Stefan Institute, Jamova 39,  SI-1000 Ljubljana, Slovenia}
\affiliation{Department of Physics,  University of Massachusetts, Amherst MA 01003, USA}

\begin{abstract}

\noindent
We present and analyze correlation functions of a main-chain polymer nematic in a continuum worm-like chain description for two types of constraints formalized by the tensorial and vectorial conservation laws,  
both originating in the microscopic chain integrity, i.e., the conectivity of the polymer chains. In particular, our aim is to identify the features of the correlation functions that are most susceptible to the differences between the two constraints.
Besides the density and director autocorrelations in both the tensorial and vectorial cases, we calculate also the density -- director correlation functions, the latter being a direct signature of the presence of a specific constraint. Its amplitude is connected to the strength of the constraint and is zero if none of the constraints is present, i.e., for a standard non-polymeric nematic.  Generally, the correlation functions  
with the constraints differ substantially from the correlation functions in the non-polymeric case, if the constraints are strong which in practice requires long chains. Moreover, for the tensorial conservation law to be well distinguishable from the vectorial one, the chain persistence length should be much smaller than the total length of the chain, so that hairpins (chain backfolding) are numerous and the polar order is small.

\end{abstract}


\maketitle

\section{Introduction}
\label{sec:intro}

\noindent
In main-chain polymer nematics the connectivity of the polymer chains forces deformations of orientational order to be inextricably linked to their density \cite{degennes,meyer,selinger,nelson0,nelson}. Splay deformations necessarily introduce local changes in polymer density, which become progressively more 
prohibitive as the chain length grows \cite{nelson0}. In the continuous limit this coupling between the polymer density and the orientational field is described by an analogue of the ``continuity equation" for the nematic director field $\bf n$ \cite{nelson,selingerJP}, that formally describes this coupling. The form of this mesoscopic continuity equation for a specific case was first suggested by de Gennes and Meyer \cite{degennes,meyer} and was later shown \cite{svensek} to depend on the detailed symmetry properties of the orientational field, i.e. whether it has a polar or quadrupolar nature,  
describable by a polar vector or a quadrupolar tensor. In these two cases the continuity equation, which is by necessity a scalar, can be cast into a  
vectorial \cite{svensek-podgornik,svensek-podgornik_chiral} or a tensorial form \cite{svensek}, respectively, and is naturally generalized to include also the vectorial/tensorial ordering moduli. The tensorial continuity equation is recent and has yet to be thoroughly investigated, ramified and implemented.
This paper presents a rather technical but indispensable step towards identifying the implications of the tensorial conservation law and preparing the grounds for detecting them in microscopic numerical simulations of generic main-chain polymer models \cite{kostas0,kostas}.

Previous theoretical considerations hinged only upon 
the vectorial constraint, irrespective of whether the underlying orientational order of the chain tangents was polar or apolar (quadrupolar), or more precisely, since the polar and quadrupolar order do not exclude each other, irrespective of the relative degree of polar and quadrupolar order of the chain tangents \cite{selinger,nelson,selingerJP,strey}. At the same time, it has been recognized and generally accepted that the polymer chain flexibility and related microscopic chain back-folding (hairpins) significantly influence macroscopic properties of nematic polymers as derived within statistical mechanics framework \cite{seitz,kamien,terentjev,edwards,linse}.

In this study we consider the consequences of microscopic chain integrity, i.e., the connectivity of monomers into a linear polymer chain, for the quadrupolar (tensorial) order parameter of a true polymer nematic which possesses the chain inversion symmetry and should thus obey the tensorial conservation law. It is hypothesized that the tensorial conservation law will eventually prevail upon the vectorial one in the limit when hairpins are abundant, i.e., when the length scales on which the tensorial constraint is applied are large compared to the persistence length of the semi-flexible main-chain polymer.
There is no other restriction in terms of a particular or chemically specific polymeric material. Moreover, since both constraints are inherent geometric/topological consequences of the unimpaired chain connectivity, they apply in principle to any physical realization of such unbreakable linear chains (e.g., also to a thread or to an elastic wire \cite{marenduzzo,hermann}) with some noise (quenched, if not thermal), described by a mesoscopic order parameter.
Our original motivation stems from the realm of biopolymers \cite{bouligand,livolant1,livolant2} described in terms of the nematic director and the density field variables, specifically including the continuum description of ordering, self-condensation, and packing of DNA in tight enclosures \cite{harvey,grason,hermann,svensek-podgornik_chiral}.

In particular, the continuity constraints certainly apply also to coarse-grained bead-spring polymer models that are convenient for computational studies \cite{marenduzzo,kostas0,kostas,skacej}.
While the limit where hairpins are abundant cannot be explored in state-of-the-art microscopic simulations, we nevertheless expect that at least significant deviations from the vectorial constraint towards the predictions of the tensorial constraint will not evade detection in the numerical simulation studies that will follow this work.

From this perspective, the primary goal of this paper is to present and analyze correlation functions of a worm-like main-chain polymer nematic constrained by the tensorial conservation law, and compare them with analogous correlation functions for the case of the vectorial conservation law \cite{nelson}. In particular, our aim is to identify the correlation functions and their features where the differences between the two constraints lead to important, possibly qualitative differences in behavior. 
With this work we are preparing a theoretical instrumentarium---a consistent, albeit idealized, prerequisite that will support and illuminate subsequent numerical studies, enabling comparisons between the theoretical correlation functions and those extracted from the simulations. Following this task, there exist a couple of essential restrictions that we intentionally adopt: 
\begin{itemize}
	\item[--] long wavelengths---while it is easier to identify qualitative differences between the two conservation laws at shorter wavelengths, only large lengths scales are relevant for comparison with simulations, as a large number of hairpins is required for manifestation of the tensorial constraint; this requirement is more severe than the usual limitation pertaining to the comparison of microscopic data with a continuum model
	\item[--] a minimal energy functional---although a description including more terms and couplings (and therewith more fitting parameters) would yield more accuracy for a specific system, it would hamper the isolation of the effects of the conservation laws. Provided that in the experimental systems (i.e., ``numerical experiments") the differences between the effects of the two constraints are expectedly small, one wants to keep the description as clean as possible in the first step
	\item[--] from the theoretical point of view, we are interested in a direct comparison of the new correlation functions calculated in the presence of the tensorial constraint, with the correlation functions for the case of the vectorial constraint, given in Refs.~\cite{selinger,nelson0,nelson}, again to identify the differences between the effects of the two constraints. Therefore we delimit ourselves to the same energy functional as in those studies.
\end{itemize}

Besides the density and director autocorrelations, we calculate also the density -- director correlation functions in both the tensorial and vectorial case. While in the context of X-ray and light scattering experiments probing the density and director fluctuations, respectively (c.f. \cite{meyer,semenovX,meyerLS,semenovLS}), these have been less relevant, nowadays they can be readily extracted from numerical simulations concurrently with the density and director autocorrelations \cite{frenkel,kostas}.  The density -- director correlation is a signature of a conservation law (either vectorial or tensorial) and its amplitude is directly connected with the strength of the constraint, being zero if none of the constraints is present, i.e., for a standard non-polymeric nematic. 


The organization of the paper is as follows. Sec.~\ref{sec:general} reviews the grounds and sets the concepts that will be used subsequently. The reader, interested mainly in the results, might skip directly to Sec.~\ref{sec:S} and forth. In Sec.~\ref{sec:general} we review the relation between microscopic and mesoscopic (coarse-grained) descriptions and elucidate their connection in Fourier space. This is important for a direct comparison of simulation/scattering data with correlation functions calculated from mesoscopic models. We review the vectorial and tensorial conservation laws. We define the correlation functions and interpret the fluctuations, extracted from microscopic degrees of freedom, in the light of fluctuating mesoscopic variables. This microscopic -- mesoscopic connection is closely related to the microscopic polymer chain connectivity and the conservation laws accounting for it on the mesoscopic level.

In Sec.~\ref{sec:F} we write down the mesoscopic free energy functional of the polymer nematic and introduce penalty potentials used to enforce the conservation laws. We discuss the hierarchy of the magnitudes of the parameters, their scaling behavior and define characteristic lengths.

In Secs.~\ref{sec:S}-\ref{sec:C} we present the correlation functions (structure factor, director autocorrelation, density -- director correlation), calculated from the free energy functionals incorporating the vectorial and tensorial constraints, respectively. We focus on the differences between the vectorial and the tensorial case. We identify the potentially measurable features of the calculated correlation functions and connect them analytically with the phenomenological parameters of the free energy functional.

\section{General premises}
\label{sec:general}

\subsection{Microscopic and coarse-grained fields}

\noindent
The following microscopic fields and their mesoscopic coarse-grained versions will be considered, where the integrals are performed over the polymer contour ${\bf x}(s)$. These quantities were introduced in Ref.~\cite{svensek}, here we list them in brief for completeness.

The microscopic segment density of the polymer chain 
\begin{equation}
	\rho^{mic}({\bf x}) =  \int_{{\bf x}(s)} ds ~\delta({\bf x} - {\bf x}(s))
	\label{rho_mic}
\end{equation}
is coarse grained to 
\begin{equation}
	\overbracket{\rho^{mic}}({\bf x}) = \rho({\bf x}) \ell_0,
	\label{rho}
\end{equation}
where $ \rho({\bf x}) = {N/V}$ is the mesoscopic number ($N$) density of monomers (or any arbitrarily defined chain segments) within the coarse-graining volume $V$ and $\ell_0$ is the corresponding segment length. 

The microscopic polymer nematic vector field 
\begin{equation}
	j_i^{mic}({\bf x}) =  \int_{{\bf x}(s)} ds~\delta({\bf x} - {\bf x}(s)) ~t_i({\bf x}(s)) ,  \qquad {\rm with} \quad t_i({\bf x}(s)) = \frac{d x_i(s)}{ds},
	\label{j_mic}
\end{equation}
is coarse grained to \cite{svensek}
\begin{equation}
	{\bf j}({\bf x}) \equiv \overbracket{{\bf j}^{mic}}({\bf x}) = \rho({\bf x})\ell_0\,{\bf a}({\bf x}),
	\label{j}
\end{equation}
where ${\bf a}({\bf x})$ is the non-unit order vector (mesoscopic-volume-average of $\bf t$) representing macroscopic polar ordering of the monomers.

The microscopic polymer nematic traceless tensor field  
\begin{equation}
	J_{ij}^{mic}({\bf x}) =  \int_{{\bf x}(s)} ds ~\delta({\bf x} - {\bf x}(s))\, 
				{\textstyle{3\over 2}}\left[t_i({\bf x}(s)) t_j({\bf x}(s))-{\textstyle{1\over 3}}\delta_{ij}\right] 
	\label{J_traceless_mic}
\end{equation}
is analogously coarse grained to \cite{svensek}
\begin{equation}
	{\sf J}({\bf x}) \equiv \overbracket{{\sf J}^{mic}}({\bf x}) = \rho({\bf x})\ell_0\,{\sf Q}({\bf x}),	
	\label{J}
\end{equation}
where $\sf Q$ is the nematic order tensor (mesoscopic-volume-average of ${\textstyle{3/2}}(t_i t_j-\delta_{ij}{\textstyle{/3}})$) representing macroscopic nematic (quadrupolar) ordering of the monomers.

One can actually express the coarse graining as a formal convolution with a coarse-graining kernel centered at each point
and having a finite extent into its neighborhood, 
which can be seen to suppress the high-$q$ components, while leaving the low-$q$ components unaltered.
This allows a direct comparison of correlation spectra acquired from raw discrete microscopic data with theoretical correlation spectra of macroscopic continuous field variables calculated from a macroscopic free energy functional, i.e., they should exactly agree in the low-$q$ region.

\subsection{Vectorial and tensorial conservation laws}

\noindent
We presented a detailed derivation of the conservation laws in the previous publication \cite{svensek}, where we showed that due to microscopic differences in the coupling between the orientational field deformations and the density variations for polar and quadrupolar order, the respective order parameters satisfy fundamentally distinct constraints. 

The mesoscopic form of the vectorial constraint valid for polar orientational order of the chain tangents is 
\begin{equation}
	\partial_i j_{i}({\bf x}) = 	\rho^+({\bf x}) - \rho^-({\bf x}),
	\label{vectorial}
\end{equation}
where ${\bf j}({\bf x}) = \rho({\bf x})\ell_0\, {\bf a}({\bf x}) $ is the mesoscopic, coarse-grained polymer current density given in Eq.~(\ref{j}),
and $\rho^\pm({\bf x})$ are the coarse-grained microscopic volume number densities of the beginnings ($\rho^+$) and the ends ($\rho^-$) of the chains defined as 
\begin{equation}
	\rho^{+mic}({\bf x}) - \rho^{-mic}({\bf x}) \equiv \delta({\bf x} - {\bf x}(0)) -  \delta({\bf x} - {\bf x}(L)),
\end{equation}
where $L$ is the length of the chain. The variable $\rho({\bf x})=\rho^+({\bf x}) - \rho^-({\bf x})$ is an additional degree of freedom describing the system.  The constraint Eq.~(\ref{vectorial}) involves also the variations of the degree of order, besides the splay deformation of the director field. In the absence of beginnings and ends of the chains, the vectorial constraint is simplified to 
\begin{eqnarray}
	\partial_i j_{i}({\bf x}) &=& 	0.
\label{vectorial0}
\end{eqnarray}

The tensorial constraint valid for quadrupolar orientational order of the chain tangents is obtained in the following mesoscopic form:  
\begin{equation}
	\partial_i\partial_j J_{ij}({\bf x})+\textstyle{1\over 2}\ell_0\nabla^2\rho({\bf x}) = 
		\nabla\cdot\left[{\bf g}^+({\bf x}) - {\bf g}^-({\bf x})\right],
		\label{tensorial}
\end{equation}
where $J_{ij}({\bf x}) = \rho({\bf x})\ell_0\, Q_{ij}({\bf x})$ is given in Eq.~(\ref{J}). 
Here ${\bf g}^\pm({\bf x})$ are the coarse-grained microscopic densities of beginnings and ends of chain tangents ${\bf t}(0), {\bf t}(L)$, defined as
\begin{equation} 
	{\bf g}^{+mic}({\bf x} ) - {\bf g}^{-mic}({\bf x} ) \equiv {\bf t}(0)\delta({\bf x} - {\bf x}(0)) - {\bf t}(L)  \delta({\bf x} - {\bf x}(L)).
\end{equation}
One can verify that $\bf g  = {\bf g}^+({\bf x})  - {\bf g}^-({\bf x})$ is invariant to interchanging heads and tails of a chain. Hence, conforming to the tensorial nature of the constraint, there is no actual need to distinguish between chain heads and tails --- they can be treated as indistinguishable, with the tangent always pointing away from the head/tail.

Generally, ${\bf g}({\bf x})$ is an additional degree of freedom of the tensorial system of polymer chains, similar to the additional variable $\rho({\bf x})$ in the vectorial case.   If the system exhibits a completely apolar orientational order, and as such does not have any local polar orientational order of the free end tangents, or if the chain beginnings/ends are absent, then ${\bf g} =0$ and Eq.~(\ref{tensorial}) reduces to 
\begin{equation}
	\partial_i\partial_j J_{ij}+\textstyle{1\over 2}\ell_0\nabla^2\rho = 0.
	\label{tensorial0}	
\end{equation}
For chains of finite length it is however expected that locally ${\bf g}$ will depart from zero self-consistently with the deformation field, analogous to a nonzero $\rho^\pm$ in the vectorial case (which is also zero in a uniform equilibrium configuration).

Just as the polar constraint in Eq.~(\ref{vectorial}) introduces a coupling between the orientational order parameter and density variation, the tensorial constraint connects the  $\sf Q$-tensor gradients with the density variations. 
Like in simulations of polymers with polar orientational order of the chain tangents \cite{svensek-podgornik,svensek-podgornik_chiral,klug}, the constraint of Eq.~(\ref{tensorial0}) would need to be enforced in coarse-grained models of nematic polymers exhibiting only quadrupolar orientational order of the chain tangents. 

For small deviations from a homogeneous configuration with quadrupolar orientational ordering of the chain tangents, the lowest order coupling between the deformations in the density and orientational fields is obtained by linearizing the constraint of Eq.~(\ref{tensorial0}) for the case where deviations from homogeneous director and density fields keep the nematic ordering uniaxial and its degree $s_0$ fixed. Assuming 
\begin{equation}
	{\bf n}=  \hat{\bf e}_z + {\delta} {\bf n}({\bf r}), \quad {\delta}{\bf n}=(\delta n_x, \delta n_y), \quad \rho = \rho_0 + \delta\rho({\bf r})
\end{equation}
and the uniaxial form $Q_{ij} = {3\over 2}s_0 (n_i n_j - {1\over 3}\delta_{ij})$ for the nematic $\sf Q$-tensor, the linearization of the tensorial constraint Eq.~(\ref{tensorial0}) leads to 
\begin{equation}
	(s_0+{\textstyle{1\over 2}})\,\partial_z^2 \delta\rho + {\textstyle{1\over 2}}(1-s_0)\, \nabla_\perp^2 \delta\rho + {\textstyle{3}} s_0\rho_0\, \partial_z \left(\nabla\!_\perp \cdot {\delta}{\bf n}\right) = 0,
	\label{linearized_tenzor}
\end{equation}
where $\nabla\!_\perp = (\partial/\partial x, \partial/\partial y)$. This is to be contrasted with the analogous linearized constraint in the vectorial case,
\begin{equation}
	a_0\,{\partial_z\delta\rho} + \rho_0 a_0\, \nabla\!_\perp\cdot{\delta}{\bf n} = 0,
	\label{linearized_vector}
\end{equation}
with ${\bf a}= a_0{\bf n}$. 
It is exactly these two reductions, Eqs.~(\ref{linearized_tenzor}) and (\ref{linearized_vector}) that can be consistently compared in the minimal description, i.e., using only the unit nematic director as the variable, which has been hitherto the exclusive case in all theoretical calculations incorporating the vectorial constraint.

\subsection{Correlations, structure factor}

\noindent
For the sake of consistency, let us briefly but explicitly introduce the (auto)correlation functions of the fluctuating variables.
Define the autocorrelation
\begin{eqnarray}
	\int d^3 x\, \langle\delta\rho^2({\bf x})\rangle 
	&=& \int\!\! d^3 x\int\!\!{d^3 q\over (2\pi)^3}\int\!\!{d^3 q'\over (2\pi)^3} \langle\delta\rho({\bf q})\delta \rho(-{\bf q}')\rangle {\rm e}^{{\rm i}({\bf q}-{\bf q}')\cdot{\bf x}}\\
	&=& \int\!\!{d^3 q\over (2\pi)^3}\,\langle\delta\rho({\bf q})\delta \rho(-{\bf q})\rangle,
\end{eqnarray}
where 
\begin{equation}
	\delta\rho({\bf q}) = \int d^3 x\, \delta\rho({\bf x}) {\rm e}^{-{\rm i}{\bf q}\cdot{\bf x}}
\end{equation}
has been defined dimensionless.
The density autocorrelation component in $\bf q$-space is known as the structure factor
\begin{equation}
	S({\bf q}) \equiv {1\over N_0}\langle \delta\rho({\bf q})\delta\rho(-{\bf q})\rangle,
	\label{S}
\end{equation}
$N_0$ is the total number of monomers.
Replacing $\delta\rho^2({\bf x})$ with other variables one gets other (auto)correlation amplitudes. In particular, we will be interested in autocorrelations of the nematic director ${\bf n}({\bf x})$,
\begin{equation}
	D_i({\bf q}) \equiv {1\over N_0}\langle\delta n_i({\bf q})\delta n_i(-{\bf q})\rangle.
	\label{D_i}
\end{equation}

Define the correlation
\begin{eqnarray}
	\int d^3 x\, \langle\delta\rho({\bf x})\delta n_i({\bf x})\rangle 
	&=& \int\!\! d^3 x\int\!\!{d^3 q\over (2\pi)^3}\int\!\!{d^3 q'\over (2\pi)^3} \langle\delta\rho({\bf q})\delta n_i(-{\bf q}')\rangle {\rm e}^{{\rm i}({\bf q}-{\bf q}')\cdot{\bf x}}\\
	&=& \int\!\!{d^3 q\over (2\pi)^3}\,\langle\delta\rho({\bf q})\delta n_i(-{\bf q})\rangle.
\end{eqnarray}
The correlation component in $\bf q$-space defined as
\begin{equation}
	C_i({\bf q}) \equiv {1\over 2N_0}\left[\langle\delta\rho({\bf q})\delta n_i(-{\bf q})\rangle + \langle\delta\rho(-{\bf q})\delta n_i({\bf q})\rangle\right]
	\label{C_i}
\end{equation}
is a real quantity. Other correlations are defined in an analogous manner.

\subsection{Extracting the (auto)correlation functions from molecular data}
\label{Sec:extracting}

\noindent
Let us address an important aspect of the connection between microscopic and mesoscopic variables, which must be taken into consideration when extracting the (auto)correlation functions of the system like those of Eqs.~(\ref{D_i}) or (\ref{C_i})  
by a scattering experiment or from a microscopic simulation as in Ref.~\cite{kostas}.

To start with, performing a discrete Fourier transform of monomer positions ${\bf x}_n$,
\begin{equation}
	\rho^{dis}({\bf q}) = \ell_0 \sum_n {\rm e}^{-i {\bf q}\cdot {\bf x}_n},
	\label{rho_dis_q}
\end{equation}
which is the discrete analogue of
\begin{equation}
	\rho^{mic}({\bf q}) = \int d^3 x\, \rho^{mic}({\bf x})\,{\rm e}^{-i {\bf q}\cdot {\bf x}}
	\label{rho_mic_q}
\end{equation}
and $\rho^{mic}$ is given by Eq.~(\ref{rho_mic}),
the low-$q$ components of the result automatically represent the low-$q$ components of the coarse-grained density $\rho\ell_0$, Eq.~(\ref{rho}). Assuming a uniform equilibrium density plus fluctuations, $\rho({\bf x})=\rho_0 + \delta\rho({\bf x})$, the equilibrium density $\rho_0$ is represented by the ${\bf q}=0$ component while the ${\bf q}\ne 0$ components are fluctuations.

In the same manner, extracting the Fourier components of the monomer tangent vectors $t_k$,
\begin{equation}
	j_k^{dis}({\bf q}) = \ell_0 \sum_n t_k\, {\rm e}^{-i {\bf q}\cdot {\bf x}_n},
	\label{j_dis_q}
\end{equation}
which is analogous to the continuum case
\begin{equation}
	j_k^{mic}({\bf q}) = \int d^3 x\, j_k^{mic}({\bf x})\,{\rm e}^{-i {\bf q}\cdot {\bf x}}
	\label{j_mic_q}
\end{equation}
defined in Eq.~(\ref{j_mic}),
the low-$q$ components represent the low-$q$ components of the coarse-grained polymer current density $\bf j$, Eq.~(\ref{j}). 
Again, the ${\bf q}=0$ component stands for the equilibrium uniform ${\bf j}_0=\rho_0\ell_0\,{\bf a}_0$, while the ${\bf q}\ne 0$ components represent fluctuations $\delta{\bf j}$. 
Note that by computing the sum of Eq.~(\ref{j_dis_q}) one extracts fluctuations of the full $\bf j$ and not merely fluctuations of $\bf a$: 
\begin{equation}
	{\delta}{\bf j}/\ell_0 = \rho_0\,{\delta}{\bf a}+{\bf a}_0\,\delta\rho = 
				\rho_0 a_0\, {\delta}{\bf n} + {\bf n}_0(\rho_0\, \delta a + a_0 \, \delta\rho),
	\label{delta_j}
\end{equation}
where we have put ${\bf a}_{(0)} = a_{(0)}{\bf n}_{(0)}$ while ${\bf n}$ and ${\bf n}_0\equiv \hat{\bf e}_z$ are unit vectors.
Due to the orthogonality of ${\delta}{\bf n}$ and ${\bf n}_0$, $\langle \delta j_i\, \delta j_i\rangle/\ell_0^2 = (\rho_0 a_0)^2\langle \delta n_i\, \delta n_i\rangle$ for $i\ne z$ nevertheless isolates the fluctuations of the preferred direction from the density/ordering fluctuations $\langle \delta j_z\, \delta j_z\rangle/\ell_0^2 = \langle [\delta(\rho a)]^2\rangle$. 
Similarly, from the correlation $\langle \delta\rho\, {\delta}{\bf j}\rangle/\ell_0 = \rho_0 a_0\langle \delta\rho\,{\delta}{\bf n} \rangle + {\bf n}_0 (\rho_0\langle \delta\rho\,\delta a \rangle + a_0\langle \delta\rho^2\rangle)$ the correlation $\langle \delta\rho\,{\delta}{\bf n} \rangle$ of the preferred direction and the density is readily obtained.

The same holds for the tensor field Eq.~(\ref{J_traceless_mic}): the extraction
\begin{equation}
	J_{kl}^{dis}({\bf q}) = \ell_0 \sum_n \textstyle{3\over 2}\left(t_k t_l-\textstyle{1\over 3}\delta_{kl}\right)\, {\rm e}^{-i {\bf q}\cdot {\bf x}_n},
	\label{J_dis_q}
\end{equation}
analogous to
\begin{equation}
	J_{kl}^{mic}({\bf q}) = \int d^3 x\, J_{kl}^{mic}({\bf x})\,{\rm e}^{-i {\bf q}\cdot {\bf x}},
	\label{J_mic_q}
\end{equation}
in the low-$q$ region yields the fluctuations of the whole ${\sf J}$, Eq.~(\ref{J}), rather than just $\sf Q$:
\begin{equation}
	{\sf\delta} {\sf J}/\ell_0 = \rho_0\,\delta {\sf Q} + {\sf Q}_0\,\delta\rho.
	\label{delta_J}
\end{equation}
Let 
\begin{equation}
	Q_{ij}=\textstyle{3\over 2} s(n_{i}n_{j}-\textstyle{1\over 3}\delta_{ij}), 	
\end{equation}
with the nematic director ${\bf n}({\bf x})$ and the equilibrium director ${\bf n}_0=\hat{\bf e}_z$ (biaxiality of both $\sf Q$ and possibly even of the equilibrium ${\sf Q}_0$ does not change the next argument).
The base tensors \cite{svensek_instab,hess} that correspond to director fluctuations, 
$({\bf\hat{e}}_z\otimes{\bf\hat{e}}_x + {\bf\hat{e}}_x\otimes{\bf\hat{e}}_z)/\sqrt{2}$
and
$({\bf\hat{e}}_z\otimes{\bf\hat{e}}_y + {\bf\hat{e}}_y\otimes{\bf\hat{e}}_z)/\sqrt{2}$
are orthogonal to ${\sf Q}_0$ and thus to the ordering and density fluctuations of $\sf J$, which are both parallel to ${\sf Q}_0$. Moreover, the director fluctuations are orthogonal also to all other types of fluctuations \cite{svensek_instab}, i.e., fluctuations of biaxiality and of a possible biaxial director.
Hence, they can be extracted from $\delta{\sf J}$ separately from all the other fluctuations. 
For $i\ne z$ we have 
\begin{eqnarray}
	\langle \delta J_{iz}\,\delta J_{iz}\rangle/\ell_0^2 &=& \left(\textstyle{3\over 2}\rho_0 s_0\right)^2\langle \delta n_i\,\delta n_i \rangle\\
	\langle \delta \rho\,\delta J_{iz}\rangle/\ell_0 &=& \left(\textstyle{3\over 2}\rho_0 s_0\right)\langle \delta\rho\,\delta n_i \rangle.
\end{eqnarray}
The role of the prefactors $\ell_0$ in Eqs.~(\ref{rho_dis_q}), (\ref{j_dis_q}), and (\ref{J_dis_q}) is merely to match the continuous versions of Eqs.~(\ref{rho_mic_q}), (\ref{j_mic_q}), (\ref{J_mic_q}). They drop out again in Eqs.~(\ref{delta_j}) and (\ref{delta_J}) when reverting to the fluctuations of the usual order parameters $\delta\bf n$ and $\delta\sf Q$.

\section{Mesoscopic free energy and the penalty potentials}
\label{sec:F}

\noindent
The correlations of Eqs.~(\ref{S}), (\ref{D_i}), and (\ref{C_i}) will be calculated using the minimal nematic free energy functional, expressed in terms of the variations of the nematic director ${\delta}{\bf n}=(\delta n_x,\delta n_y)$ and the density $\delta\rho$. The $(x,y)$ subspace is denoted by $\perp$. As before, the equilibrium director is along $z$ and the equilibrium density is denoted $\rho_0$.
The density and nematic part of the free energy reads\begin{equation}
	{\cal F}_{\rho n} = {1\over 2}\int\!\!\!\!\int\! {d}z\,{d}^2r_\perp
	    \left[B\left(\delta\rho\over\rho_0\right)^2 + B'\left({\vert\nabla\delta\rho\vert\over\rho_0}\right)^2 +
	    	  K_1\left(\nabla\!_\perp\cdot{\delta}{\bf n}\right)^2 + K_2\left(\nabla\!_\perp\times{\delta}{\bf n}\right)^2 +
	    	  K_3\vert\partial_z{\delta}{\bf n}\vert^2\right],
	\label{Fnem}
\end{equation}
where $K_1$, $K_2$, and $K_3$ are the Frank elastic constants for splay, twist, and bend. 
The last term in fact embodies the continuum version of the worm-like chain model as discussed in Ref. \cite{strey}.
A term ($B'$) quadratic in the density gradient has been included solely for the purpose of defining a microscopic length scale. 
In principle, this term should be also reflecting the nematic symmetry and distinguish between $\partial/\partial z$ and $\nabla_\perp$. 
In any case, the $B'$ term would become significant only at the length scale defined by the density correlation length $\xi_\rho\sim \sqrt{B'/B}$ and smaller, which we will stay strictly away from, as discussed in Sec.~\ref{sec:intro}.
It was included in the free energy functional only as an indicator, reminding one to concentrate on length scales where the effect of $B'$ in the correlation functions is not noticeable, c.f. Eqs.~(\ref{vqz})-(\ref{tqp}).

The conservation laws will be enforced by penalty potentials to be added to the functional Eq. (\ref{Fnem}).
The penalty potential imposing the vectorial constraint Eq. (\ref{vectorial0}) under the assumption of fixed dipolar orientational order $a_0$ of the chain tangents reads
\begin{equation}
	{\cal F}^{v} = {1\over 2}G\ell_0^2 a_0^2\int\!\!\!\!\int\! {d}z\,{d}^2 r_\perp\,
	\left(\partial_z\delta\rho + \rho_0 \nabla\!_\perp\cdot{\delta}{\bf n}\right)^2, 
	\label{F_v}
\end{equation}
where $G$ is the phenomenological strength of the constraint, the value of which can be derived from a simple microscopic model treating the chain heads and tails as an ideal gas \cite{nelson}, and the square contains the linearized form of the vectorial constraint, Eq.~(\ref{linearized_vector}).

The penalty potential imposing the tensorial constraint, Eq. (\ref{tensorial0}), for the equilibrium ${\sf Q}_{0}=\textstyle{3\over 2} s_0(\hat{\bf e}_z\otimes\hat{\bf e}_z-\textstyle{1\over 3}{\sf I})$ and director fluctuations only, reads
\begin{equation}
	{\cal F}^{t} = {1\over 2}H\ell_0^2 \int\!\!\!\!\int\! {d}z\,{d}^2 r_\perp\,
	\left[
	(s_0+{\textstyle{1\over 2}})\,\partial_z^2 \delta\rho + {\textstyle{1\over 2}}(1-s_0)\, \nabla_\perp^2 \delta\rho + {\textstyle{3}} s_0\rho_0\, \partial_z \left(\nabla\!_\perp \cdot {\delta}{\bf n}\right)
	\right]^2,
	\label{F_t}
\end{equation}
where $H$ is again the phenomenological strength of the tensorial constraint and the square now contains the linearized form of the tensorial constraint, Eq.~(\ref{linearized_tenzor}).

In the functional Eq.~(\ref{Fnem}), symmetry would allow also for two other terms that couple gradients of the director and gradients of the density \cite{HH}: 
$(\nabla\!_\perp\cdot{\delta}{\bf n})(\partial_z\rho)$ and
$(\partial_z {\delta}{\bf n})\cdot (\nabla\!_\perp \rho)$,
which had not been included in the minimal free energy of Refs.~\cite{selinger,nelson} that we have adopted as a point of departure. While in principle these terms describe additional effects not covered by our 
functional, we nevertheless omit them here, arguing that they will be eventually made superfluous by the constraints whose strength increases directly in proportion to the polymer chain length. 
Nevertheless, we do point out that in particular the second coupling introduces a new effect (coupling of director bend deformation and density variations) that might play a role in a future more refined analysis, when such a refinement will be justified by additional evidence.


Moreover, in the spirit of the minimal description pursued here we will also not consider spatial variations of the degree of nematic ordering and the resulting cross-coupling terms \cite{kawasaki} between its gradients and gradients of the director, leaving it for future refinements. In fact, inclusion of the moduli fluctuations is the next natural extension of our present treatment: in principle they should be taken into account even in the large length scale limit pursued here (where variations of the ordering moduli are normally irrelevant) and even if the cross-couplings themselves were inferior. Namely, variations of the moduli are dictated by the constraints on equal basis as variations of the density, and there is no a priori reason why the latter should be more important than the former. In the vectorial constraint they appear in analogous forms. The tensorial constraint, in contrast, being second order in the gradients itself introduces a coupling between both types of variations.
To establish an one-to-one comparison of the two conservation laws entirely within the minimal director picture, we neglect the variations of the ordering moduli in this first approach.

We remind of the discussion in Sec.~\ref{Sec:extracting}, pointing out that the director fluctuations are orthogonal to the fluctuations of the nematic ordering and the biaxiality. 
This means that they can be separately extracted from the microscopic data.
The assumptions of constant $s_0$ and zero biaxiality in Eq.~(\ref{F_t}) as well as constant $a_0$ in Eq.~(\ref{F_v}) are however approximations.

\begin{figure}
\begin{center}
	\includegraphics[width=7cm]{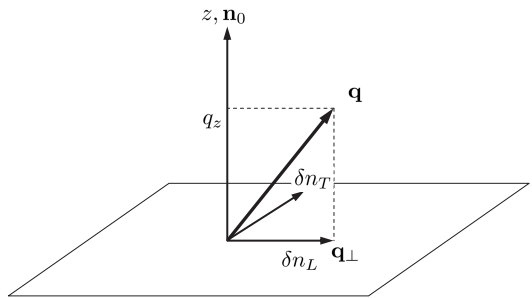}
	\caption{Definition of the directions, ${\bf q} = ({\bf q_\perp},q_z), {\delta}{\bf n}=(\delta n_L,\delta n_T,0)$; $\delta n_L$ is the component along $\bf q_\perp$. The system is symmetric about the $z$ axis, i.e., the direction of $\bf q_\perp$ in $xy$ plane is chosen arbitrarily.}
\label{fig:directions}
\end{center}
\end{figure}

Our mesoscopic free energy functionals are thus ${\cal F}_{\rho n}^v={\cal F}_{\rho n}+{\cal F}^v$ for the case of the vectorial constraint and ${\cal F}_{\rho n}^t={\cal F}_{\rho n}+{\cal F}^t$ for the case of the tensorial constraint.
In $\bf q$-space, for a given ${\bf q}=({\bf q}_\perp,q_z)$, thus for the variables $\delta\rho({\bf q})$ and  ${\delta}{\bf n}({\bf q})=(\delta n_L({\bf q}),\delta n_T({\bf q}))$ expressed in longitudinal and transversal components with respect to ${\bf q}_\perp$, depicted in Fig.~\ref{fig:directions}, the density and nematic part is 
\begin{equation}
	f_{\rho n}({\bf q}) = {1\over 2}\left[\tilde{B}\left(\delta\rho\over\rho_0\right)^2 + 
		(K_1 q_\perp^2 + K_3 q_z^2)\, \delta n_L^2 + 
		(K_2 q_\perp^2 + K_3 q_z^2)\, \delta n_T^2 \right],
	\label{f_q}
\end{equation}
where $\tilde{B} = B+B' q^2$ or, more precisely for the nematic fluid, $\tilde{B} = B + B'_\parallel q_z^2 + B'_\perp q_\perp^2$, 
and the complete vectorial and tensorial functionals are
\begin{eqnarray}
	f_{\rho n}^v({\bf q}) &=& f_{\rho n}({\bf q}) + 
		{1\over 2} G\ell_0^2\,a_0^2\left(q_z\,\delta\rho + \rho_0\, q_\perp\,\delta n_L \right)^2,\label{fv_q}\\
	f_{\rho n}^t({\bf q}) &=& f_{\rho n}({\bf q}) +
		{1\over 2} H\ell_0^2\left\{\left[(s_0+{\textstyle{1\over 2}})q_z^2+{\textstyle{1\over 2}}(1-s_0)q_\perp^2\right]\delta\rho + \textstyle{3}\rho_0 s_0\, q_z q_\perp\,\delta n_L \right\}^2.\label{ft_q}	
\end{eqnarray}
For brevity of notation, in Eqs.~(\ref{f_q})--(\ref{ft_q}) we have replaced $u u^*$ with $u^2$, where $u$ stands for any of the variables $\delta\rho$, $\delta n_L$, $\delta n_T$, or their combinations.
The free energy is ${\cal F} = \int {d^3 q/(2\pi)^3}\, f({\bf q})$.

\subsection{Parameter magnitudes hierarchy and scaling}

\noindent
We will use $B$ as the reference parameter. The ratio $\sqrt{K/B}\equiv \xi$ defines a microscopic length scale, at which director deformations and density variations are equally costly. Using $\xi$ as the length unit, $K=B$ and the low-$q$ region corresponds to $q\ll 1$. In this long wavelength regime, director deformations are cheap compared to density variations and also $\tilde{B}\approx B={\rm const}$.

Let us define $\tilde{G}\equiv G\ell_0^2\rho_0^2 a_0^2$ and $\tilde{H}\equiv H\ell_0^2\rho_0^2$.
The strength of the constraints is controlled by the ratios $\tilde{G}/B$ and $\tilde{H}/B$, respectively. It follows from 
the comparison of the ${\bf q}$-dependence of the penalty potentials in Eqs.~(\ref{fv_q})-(\ref{ft_q}) with respect to $f_{\rho n}^v({\bf q})$ and $f_{\rho n}^t({\bf q})$, respectively, 
that the effective strength of the constraints depends on the length scale of the deformation. The vectorial constraint is rigid for $1/q^2\ll \tilde{G}/B$ and vanishes for $1/q^2\gg \tilde{G}/B$. The tensorial constraint is rigid for $1/q^4\ll \tilde{H}/B$ and vanishes for $1/q^4\gg \tilde{H}/B$. For the constraints to be effective in the $q\ll 1$ region, we must have $\tilde{G}/B\gg 1$ and in particular $\tilde{H}/B\gg 1$.

In the next Sections we aim to calculate and analyze the correlation functions for the case of the tensorial constraint based on the functional Eq.~(\ref{ft_q}). We compare them to the correlation functions for the case of the vectorial constraint \cite{nelson} based on the functional Eq.~(\ref{fv_q}), which we also re-derive along the same lines. The correlation functions will be presented in units of $k_B T \rho_0^n/B$, where $n=\{1,0,-1\}$ for $\{S({\bf q}),C({\bf q}),D({\bf q})\}$, respectively. In these units, $B=1$. In all plots we will use $B'_\parallel=B'_\perp=1$.

\section{The structure factor, 	$S({\bf q})$}

\label{sec:S}

\noindent
As an introduction and verification of our procedure, we first reproduce the structure factor, Eq.~(\ref{S}), given in Eq.~(5.5) of Ref.~\cite{nelson}, for the nematic vector order parameter ${\bf a}=a_0{\bf n}$, where $\bf n$ is the unit vector and the degree of order $a_0$ is fixed. It is important to realize that in this limit, $\bf n$ can as well represent (and usually does) the nematic director, i.e., the axis of the (uniaxial) nematic order tensor.
The director formulation has been standardly used for the nematic (quadrupolar) ordering, while at the same time using the vectorial conservation law Eq.~(\ref{vectorial}).

\begin{figure}
\begin{center}
	\mbox{
	\includegraphics[width=5.5cm]{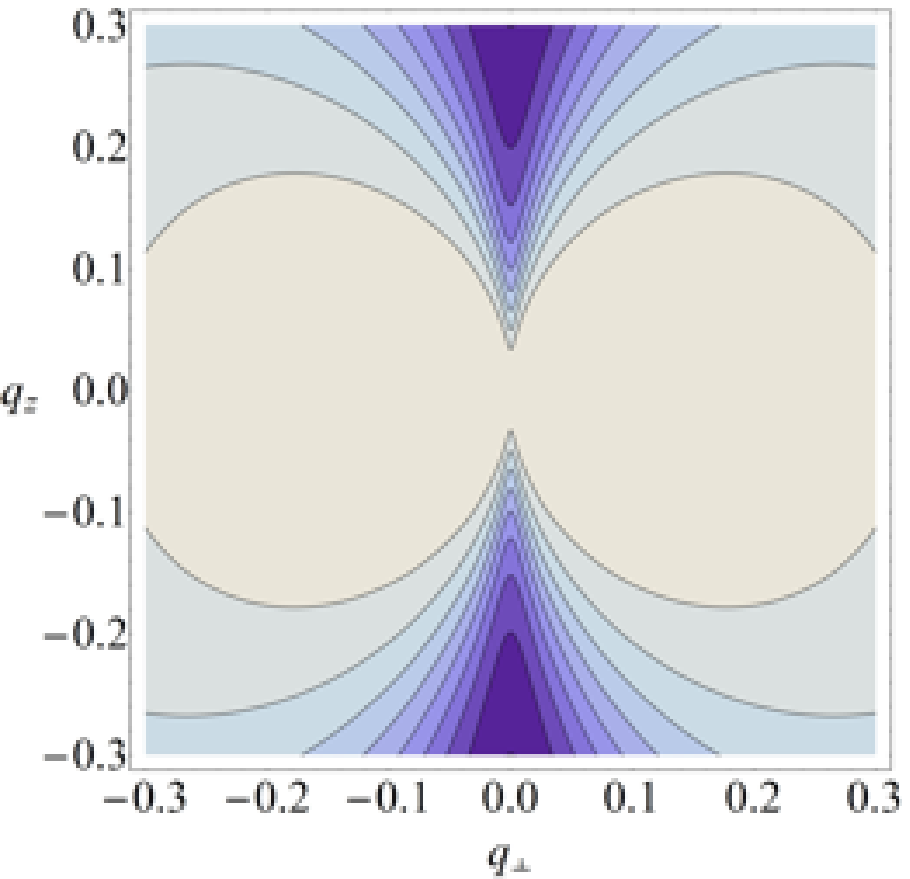}\hspace{1mm}
	\includegraphics[width=6cm]{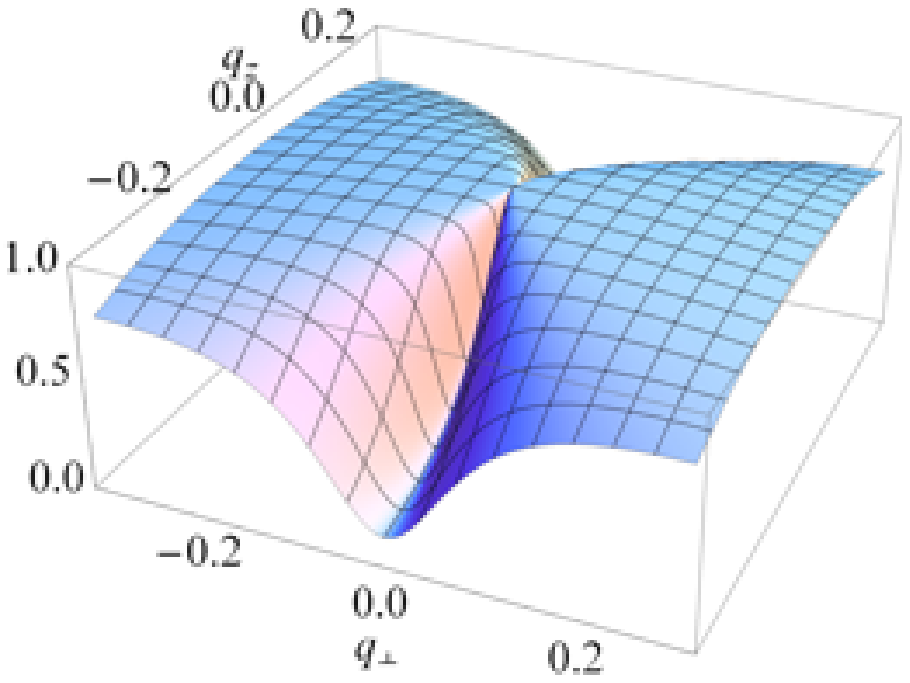}
	}
	\caption{(color online) The vectorial structure factor $S^v(q_\perp, q_z)$ of Eq.~(\ref{SvectG}) in the low-$q$ region: density fluctuations in the directions around $q_\perp=0$ are suppressed. $K_1=0.5$, $K_3=1$, $\tilde{G}=100$.}
\label{fig:Svect}
\end{center}
\end{figure}

The structure factor following from the density--nematic free energy functional with the vectorial constraint, Eq.~(\ref{fv_q}), is
\begin{equation}
	S^v({\bf q}) = 
	k_B T \rho_0{q_\perp^2 + {\left(K_1 q_\perp^2+K_3 q_z^2\right)/\tilde{G}}\over
					 \tilde{B} q_\perp^2 + \left({\tilde{B}/\tilde{G}}+q_z^2\right)\left(K_1 q_\perp^2+K_3 q_z^2\right)}
	\label{SvectG}
\end{equation}
and is shown in Fig.~\ref{fig:Svect}.
It does not involve the twist elastic constant $K_2$ as the transversal director component $\delta n_T$ is not coupled to the density.
The structure factor Eq.~(\ref{SvectG}) should be compared to the structure factor of a non-polymeric nematic fluid ($G=0$),
\begin{equation}
	S({\bf q}) = k_B T {\rho_0\over\tilde{B}},
	\label{S_noG}
\end{equation}
recalling that also in this case $\tilde{B} = B + B'_\parallel q_z^2 + B'_\perp q_\perp^2$. 
We note that we are dealing with the volume number density of monomers, in contrast to Ref.~\cite{nelson}, where the areal density of chains is considered. 


In the case of the tensorial conservation law Eq.~(\ref{tensorial}), the structure factor, assuming only director fluctuations of the nematic $\sf Q$-tensor, follows from the functional Eq.~(\ref{ft_q}) enforcing the tensorial constraint:
\begin{equation}
	S^t({\bf q}) = 
	k_B T\rho_0 {(3 s_0\, q_z q_\perp)^2 + \left(K_1 q_\perp^2+K_3 q_z^2\right)/\tilde{H}\over
	       \tilde{B} (3s_0\, q_z q_\perp)^2 + \left\{{\tilde{B}/\tilde{H}}+\left[(s_0+{\textstyle{1\over 2}})q_z^2+
	       							       {\textstyle{1\over 2}}(1-s_0)q_\perp^2\right]^2\right\}
	       				\left(K_1 q_\perp^2+K_3 q_z^2\right)}.
	\label{StensG}
\end{equation}
Besides the $s_0$-dependence coming from the tensorial constraint, Eq. (\ref{F_t}), which is shown explicitly in Eq.~(\ref{StensG}), the elastic constants depend on the degree of order as well, and $K_i\propto s_0^2$ to the lowest order as usual \cite{schiele}. From the viewpoint of the tensorial conservation law, the interesting part of the $s_0$-dependence is near $s_0=1$ and comes from Eq.~(\ref{F_t}).

\begin{figure}
\begin{center}
	\mbox{
	\includegraphics[width=5.5cm]{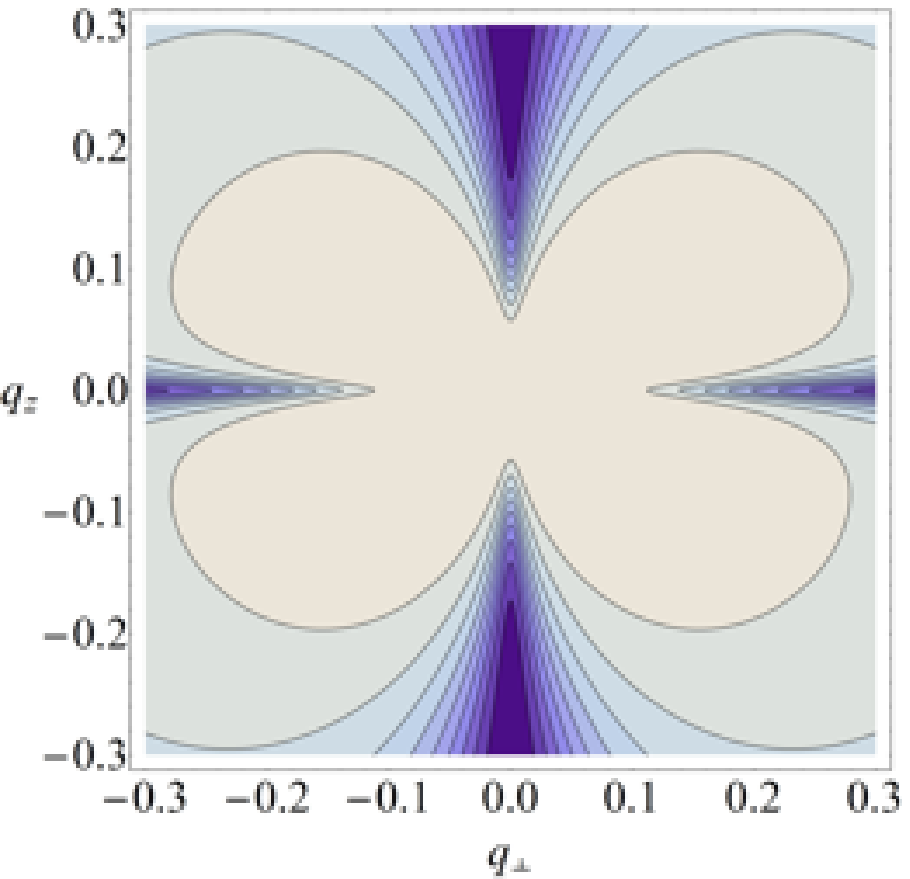}\hspace{1mm}
	\includegraphics[width=6cm]{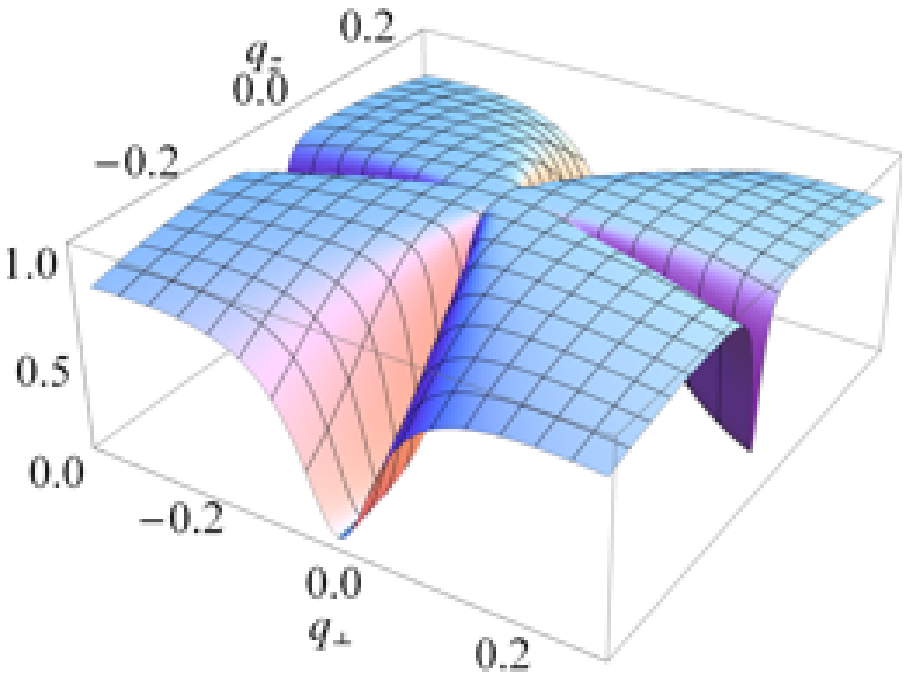}
	}
	\caption{(color online) The tensorial structure factor $S^t(q_\perp, q_z)$ of Eq.~(\ref{StensG}) in the low-$q$ region: also suppressed are the density fluctuations in the directions around $q_z=0$. $K_1=0.5$, $K_3=1$, $\tilde{H}=10000$, $s_0=0.5$.}
\label{fig:Stens}
\end{center}
\end{figure}

Fig.~\ref{fig:Stens} shows that in this case density fluctuations are suppressed also in the directions close to $q_z=0$, if only $s_0<1$.  
One can verify that the non-polymeric limit $H=0$ is again given by Eq.~(\ref{S_noG})
and the non-polymeric structure factor 
in the low-$q$ region presented in Figs.~\ref{fig:Svect} and \ref{fig:Stens} by comparison is essentially flat.


\subsubsection{Wave vector parallel to the director ($q_\perp=0$)}
\label{sec:parallel}

\noindent
The vectorial case, Eq.~(\ref{SvectG}), becomes
\begin{equation}
	S^v(0,q_z) = 
	k_B T \rho_0 {1 \over
		B+(B' + \tilde{G}) q_z^2},
	\label{vqz}
\end{equation}
i.e., a parabolic $q_z$-dependence of $1/S^v$; $B'$ is negligible.
The tensorial case, Eq.~(\ref{StensG}), becomes
\begin{equation}
	S^t(0,q_z) = 
	k_B T \rho_0 {1 \over
		B+B'q_z^2 + \tilde{H}(s_0+\textstyle{1\over 2})^2 q_z^4},
	\label{tqz}
\end{equation}
where $B'$ is negligible except at the very origin. Thus, this time we observe a quartic $q_z$-dependence of $1/S^t$.

\subsubsection{Wave vector perpendicular to the director ($q_z=0$)}
\label{sec:perpendicular}

\noindent
The vectorial case, Eq.~(\ref{SvectG}), becomes
\begin{equation}
	S^v({\bf q}_\perp,0) = 
	k_B T \rho_0 {1 + K_1/\tilde{G}\over
		(B+B'q_\perp^2)\left(1 + K_1/\tilde{G}\right)},
	\label{vqp}
\end{equation}
which is essentially a constant.
The tensorial case, Eq.~(\ref{StensG}), on the other hand becomes
\begin{equation}
	S^t({\bf q}_\perp,0) = 
	k_B T \rho_0 {1 \over
		(B+B'q_\perp^2) + {\textstyle 1\over 4}\tilde{H}(1-s_0)^2 q_\perp^4},
	\label{tqp}
\end{equation}
which if $s_0<1$ gives a quartic $q_\perp$-dependence of $1/S^t$.

\section{Director autocorrelation, $D_L({\bf q})$}
\label{sec:D}

\noindent
Here we calculate the director autocorrelation functions Eq.~(\ref{D_i}).
The transversal component $\delta n_T$ is decoupled and thus
\begin{equation}
	\langle \delta n_T({\bf q})\delta n_L(-{\bf q})\rangle = 0
\end{equation}
and
\begin{equation}
	D_T({\bf q}) = {k_B T\over\rho_0} {1\over K_2 q_\perp^2 + K_3 q_z^2}
\end{equation}
like in a regular nematic. This is valid for both the vectorial and tensorial cases.

\begin{figure}
\begin{center}
	\mbox{
	\includegraphics[width=5.5cm]{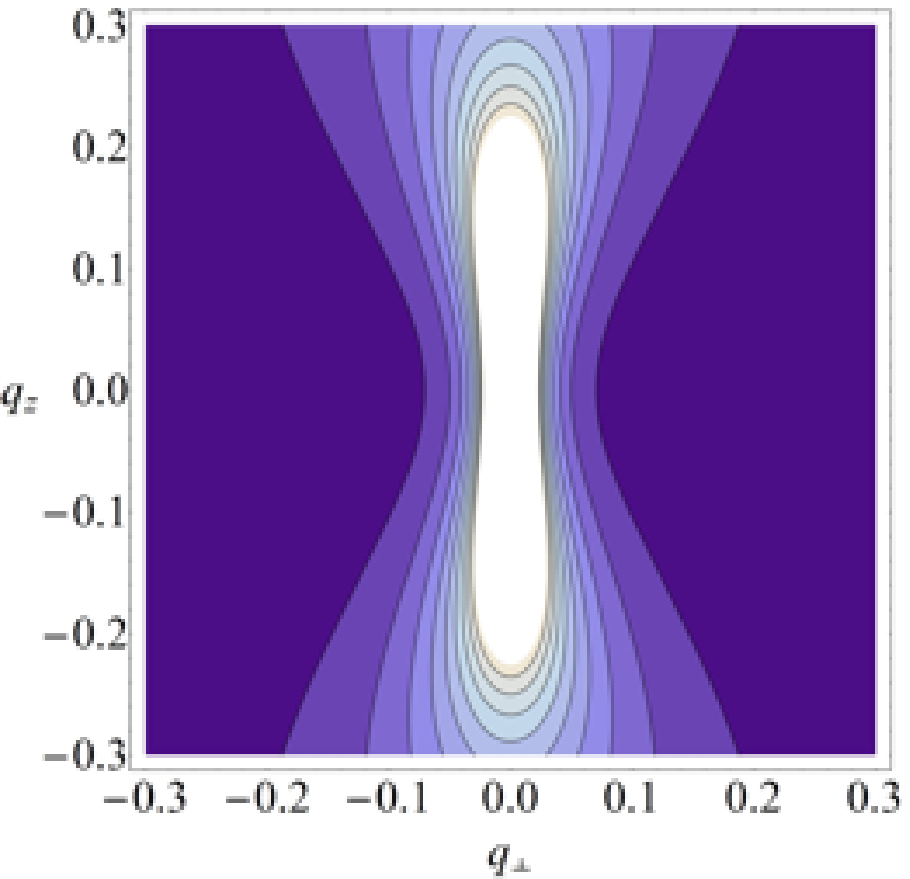}\hspace{1mm}
	\includegraphics[width=6cm]{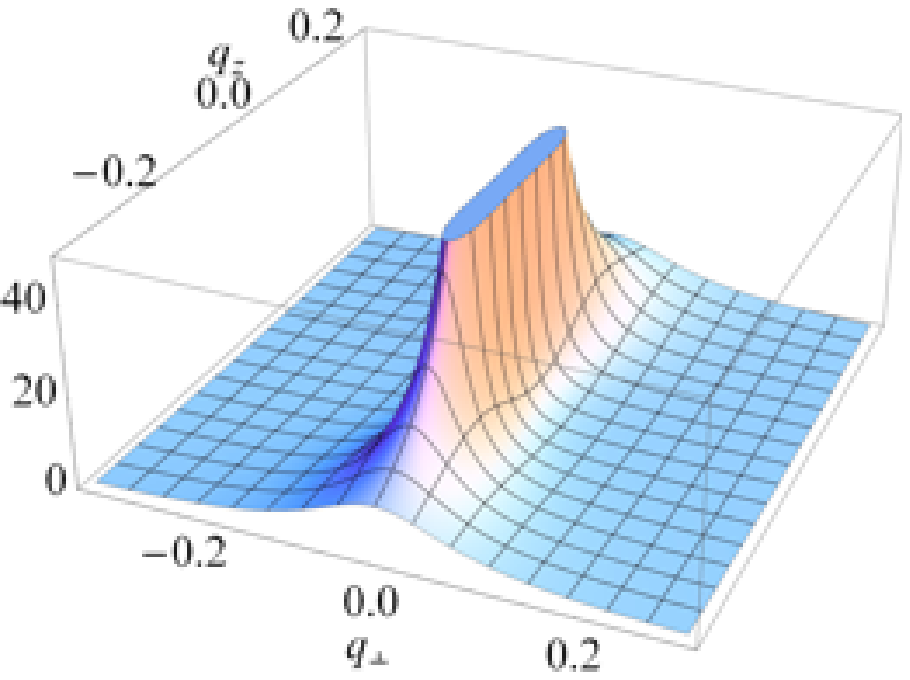}
	}
	\caption{(color online) Vectorially constrained director autocorrelation function $D_L^v(q_\perp, q_z)$ of Eq.~(\ref{D_Lv}) in the low-$q$ region: director fluctuations with $q_\perp\ne 0$ are strongly suppressed. $K_1=0.5$, $K_3=1$, $\tilde{G}=100$.}
\label{fig:Dvect}
\end{center}
\end{figure}
\begin{figure}
\begin{center}
	\includegraphics[width=6cm]{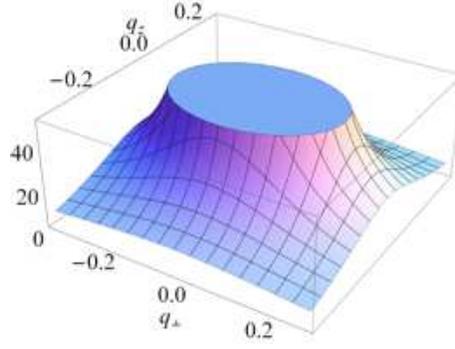}
	\caption{(color online) Director autocorrelation function of the non-polymeric nematic $D_L(q_\perp, q_z)$ of Eq.~(\ref{D_Lnem}) in the same region and scale as Figs.~\ref{fig:Dvect} and \ref{fig:Dtens}. $K_1=0.5$, $K_3=1$.}
\label{fig:DnoG}
\end{center}
\end{figure}
\begin{figure}
\begin{center}
	\mbox{
	\includegraphics[width=5.5cm]{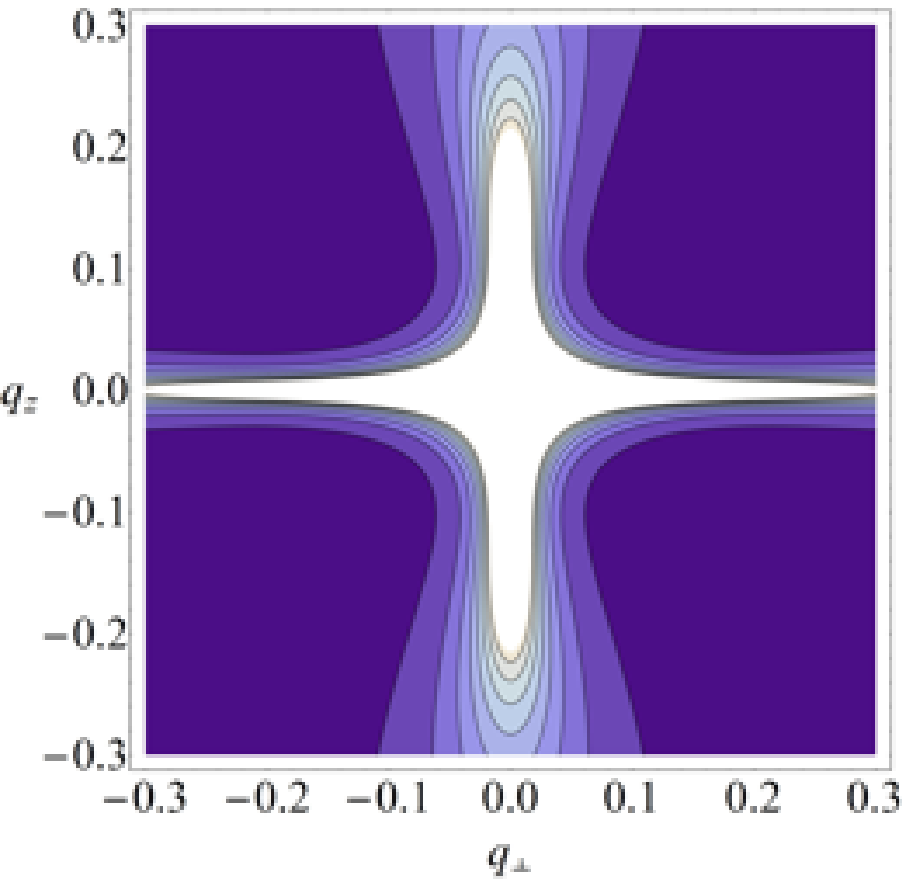}\hspace{1mm}
	\includegraphics[width=6cm]{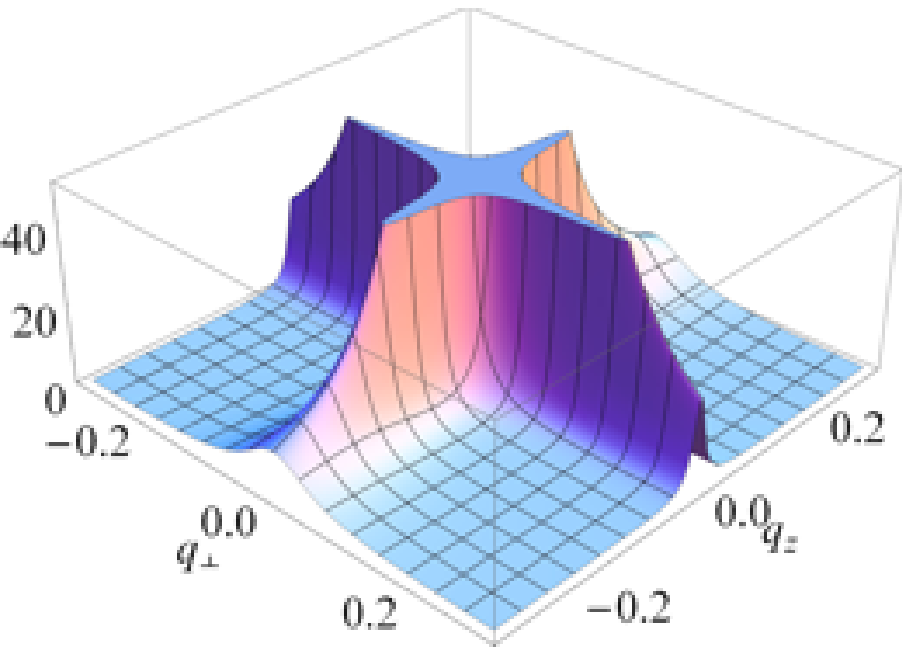}
	}
	\caption{(color online) Tensorially constrained director autocorrelation function $D_L^t(q_\perp, q_z)$ of Eq.~(\ref{D_Lt}) in the low-$q$ region: director fluctuations with oblique direction with both $q_\perp$ and $q_z$ nonzero are strongly suppressed. $K_1=0.5$, $K_3=1$, $\tilde{H}=10000$, $s_0=0.5$.}
\label{fig:Dtens}
\end{center}
\end{figure}

In the vectorial case, the autocorrelation of the longitudinal director component, following from the functional Eq.~(\ref{fv_q}), is
\begin{equation}
	D_L^v({\bf q}) = {k_B T\over\rho_0} {\tilde{B}/\tilde{G} + q_z^2\over 
		\left(\tilde{B}/\tilde{G} + q_z^2\right)\left(K_1 q_\perp^2+K_3 q_z^2\right)+\tilde{B}q_\perp^2}
	\label{D_Lv}
\end{equation}
and is shown in Fig.~\ref{fig:Dvect}.
In the non-polymeric limit ($G=0$) the result for the regular nematic is recovered,
\begin{equation}
	D_L({\bf q}) = {k_B T\over\rho_0}{1\over K_1 q_\perp^2 + K_3 q_z^2},
	\label{D_Lnem}
\end{equation}
which is plotted in Fig.~\ref{fig:DnoG} for comparison with Figs.~\ref{fig:Dvect} and \ref{fig:Dtens}.
Using Eq.~(\ref{D_Lv}) and putting $q_z=0$, by comparison with Eq.~(\ref{D_Lnem}) one can identify, similar to Ref.~\cite{nelson}, a renormalized splay constant $K_1^R$:
\begin{equation}
	D_L^v({\bf q}_\perp,0) = {k_B T\over\rho_0} {1\over K_1^R\,q_\perp^2},\quad
	K_1^R = K_1 + \tilde{G},
\end{equation}
which should be a linear function of the chain length, according to the ideal gas model of the chain ends \cite{nelson}.


In the tensorial case, the autocorrelation of the longitudinal director component, following from the functional Eq.~(\ref{ft_q}), is
\begin{equation}
	D_L^t({\bf q}) = {k_B T\over\rho_0} {{\tilde{B}/\tilde{H}}+\left[(s_0+\textstyle{1\over 2})q_z^2+\textstyle{1\over 2}(1-s_0)q_\perp^2\right]^2\over 
	\left\{{\tilde{B}/\tilde{H}}+\left[(s_0+\textstyle{1\over 2})q_z^2+\textstyle{1\over 2}(1-s_0)q_\perp^2\right]^2\right\}\left(K_1 q_\perp^2+K_3 q_z^2\right)+\tilde{B}(3s_0\, q_z q_\perp)^2}
	\label{D_Lt}
\end{equation}
and is shown in Fig.~\ref{fig:Dtens}.

\section{The density -- director correlation, $C_L({\bf q})$}
\label{sec:C}

\noindent
The density -- director correlation function Eq.~(\ref{C_i}) is particularly interesting, since it is a direct signature of the polymer conservation laws and is zero in a non-polymeric nematic. Moreover, the peaks and other characteristic features of these correlation functions depend on the strengths of the constraints $\tilde{G}$ and $\tilde{H}$ and thus trail the extent to which these constraints are imposed. 

The transversal director component is not correlated to the density (no coupling in the free energy), $C_T({\bf q})=0$. On the other hand, it can be shown that the nontrivial correlation $C_L({\bf q})$ is connected to $D_L({\bf q})$.
Due to the fact that the functions $C_L^v({\bf q})$ and $C_L^t({\bf q})$ directly reflect the vectorial and tensorial constraints, respectively, we perform a closer analysis of their $\bf q$-space landscapes presented in Figs.~\ref{fig:Cvect} and \ref{fig:Ctens}.

\subsection{Vectorial case, $C_L^v({\bf q})$}

\begin{figure}
\begin{center}
	\mbox{
	\includegraphics[width=5.5cm]{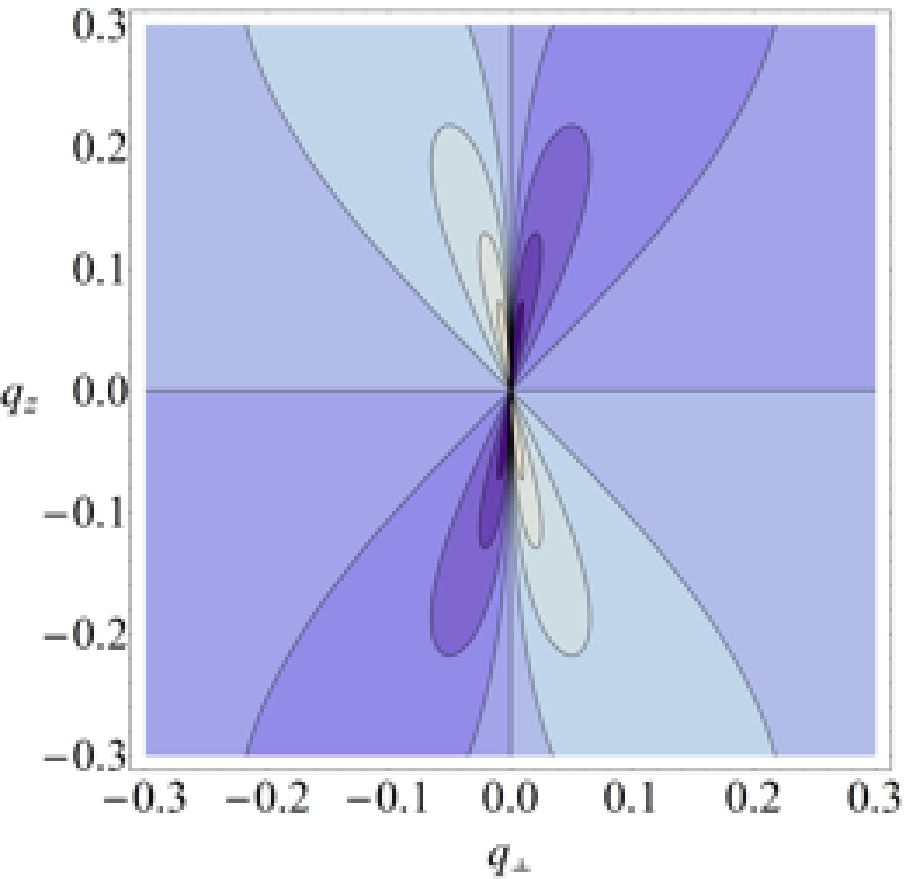}\hspace{1mm}
	\includegraphics[width=6cm]{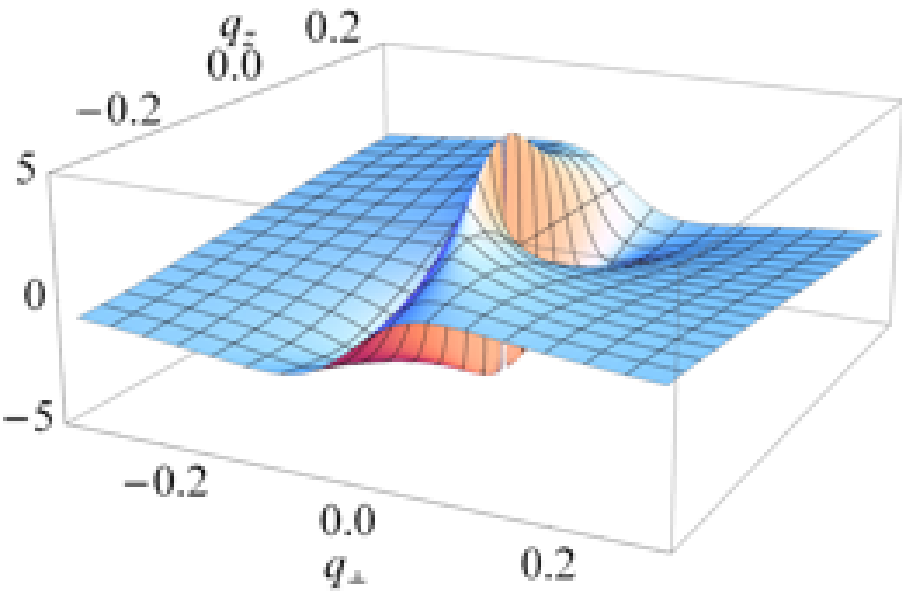}
	}
	\caption{(color online) Density -- director correlation due to the vectorial constraint $C_L^v(q_\perp, q_z)$ of Eq.~(\ref{C_Lv}) in the low-$q$ region. $K_1=0.5$, $K_3=1$, $\tilde{G}=100$.}
\label{fig:Cvect}
\end{center}
\end{figure}

\begin{figure}
\begin{center}
	\mbox{
	\subfigure[]{\includegraphics[height=6cm]{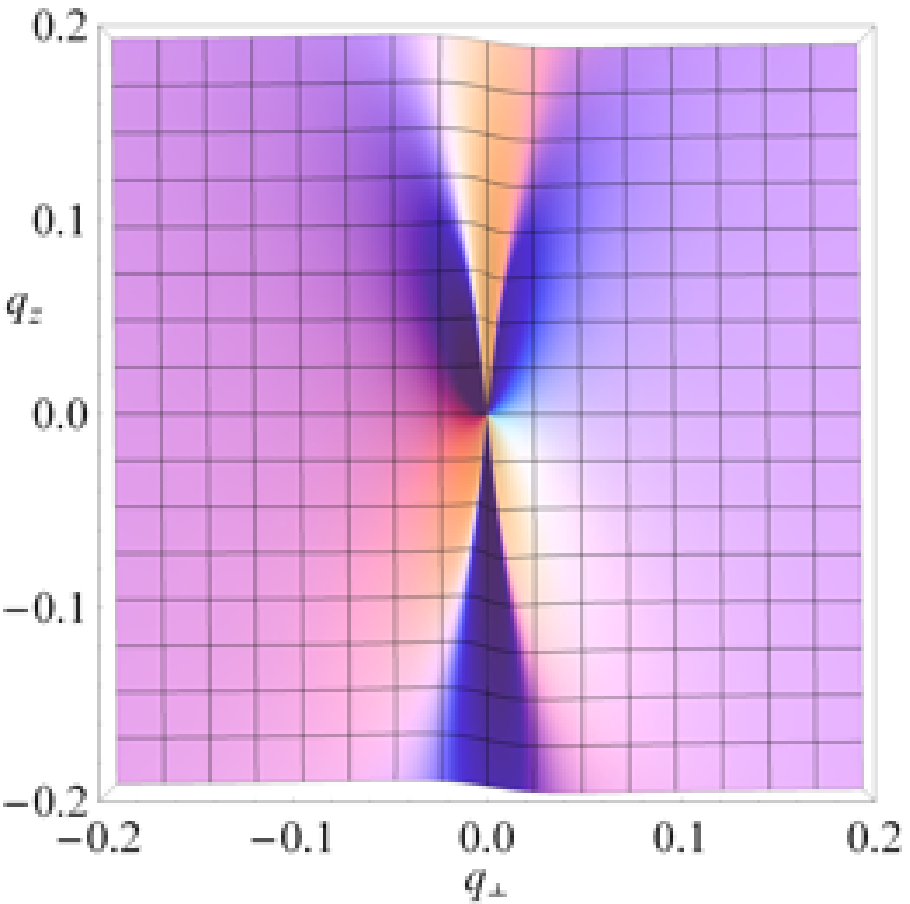}
	}\hspace{5mm}
	\subfigure[]{\includegraphics[height=6cm]{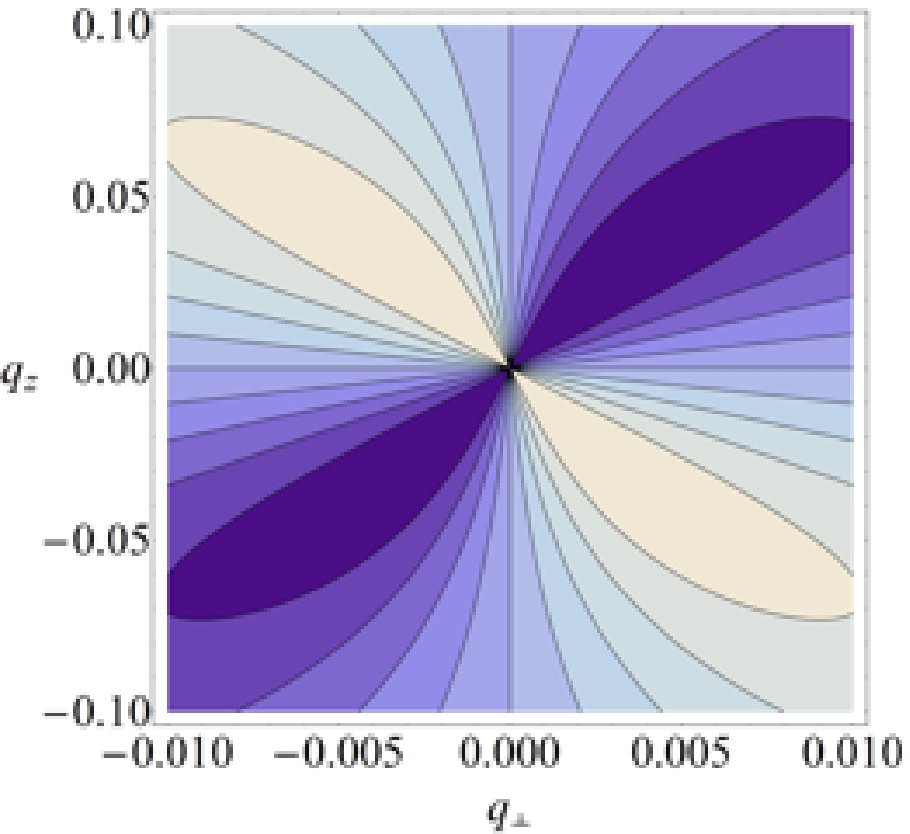}
	}
	}
	\caption{(color online) (a) Top view of the vectorial density -- director correlation landscape, revealing linear (for $q_z\to 0$) and parabolic dependence of $q_\perp(q_z)$ of the crests. (b) The contours of constant $C_L^v$ closer to the origin (note the 10$\times$ magnification in the $q_\perp$ direction). $K_1=0.5$, $K_3=1$, $\tilde{G}=100$.}
\label{fig:Cvect_details}
\end{center}
\end{figure}

\noindent
In the vectorial case, the connection between $C_L^v({\bf q})$ and $D_L^v({\bf q})$ is 
\begin{equation}
	C_L^v({\bf q}) =-\rho_0{q_z q_\perp\over\tilde{B}/\tilde{G} + q_z^2}\, D_L^v({\bf q})
\end{equation}
and thus with Eq.~(\ref{D_Lv}) the density -- director correlation due to the vectorial constraint is
\begin{equation}
	C_L^v({\bf q}) = -k_B T {q_z q_\perp\over \left({\tilde{B}/\tilde{G}}+q_z^2\right)\left(K_1 q_\perp^2+K_3 q_z^2\right)+\tilde{B}q_\perp^2}.
	\label{C_Lv}
\end{equation}
Note that the correlation is zero for $G=0$ as mentioned before. The function $C_L^v(q_\perp,q_z)$ is presented in Fig.~\ref{fig:Cvect}, showing narrow ridges in the $q_z$ direction close to the $q_\perp=0$ axis, peaking at $q=0$. One can show that
\begin{equation}
	C_L^v(q\to 0) \approx \pm {k_B T\over B} {1\over 2}\sqrt{\tilde{G}\over K_3},
\end{equation}
i.e., the maximum value of the correlation is controlled by the ratio $\tilde{G}/K_3$.
Further it can be shown that the points at the top/bottom of the ridges are related by
\begin{equation}
	q_\perp = \sqrt{(B/\tilde{G}+q_z^2)K_3\over (B/\tilde{G}+q_z^2)K_1 + B}\, q_z,
\end{equation}
resulting in two regimes: straight lines $q_\perp \approx \pm\sqrt{K_3/\tilde{G}}\, q_z$ for $q_z^2\ll {B/\tilde{G}}$ and parabolas $q_\perp \approx \pm\sqrt{K_3/B}\, q_z^2$ for $q_z^2\gg {B/\tilde{G}}$ while still $K_1 q_z^2\ll B$. (The latter two conditions can be met simultaneously since $\tilde{G}\gg K_1$, usually). Both features can be traced on Fig.~\ref{fig:Cvect_details}a, showing the ridges from the top.

In the parabolic regions the height of the ridge decays as
\begin{equation}
	C_L^v(q_\perp=\sqrt{\textstyle{K_3\over B}}q_z^2,q_z) \approx \pm {k_B T\over B}{1\over 2}\sqrt{B\over K_3}\,{1\over q_z}.
\end{equation}

For $q_z\ll B/\tilde{G}$, the contours of constant $C_L^v$ are straight lines with zero intercept and slope satisfying
\begin{equation}
	{q_\perp\over q_z} \approx {1\over C_L^v B}\left[\left(K_3{q_z^2\over q_\perp^2}+K_1\right)/\tilde{G}+1\right]^{-1},
\end{equation}
which defines two sets of lines, one at either side of the ridges, Fig.~\ref{fig:Cvect_details}b. 
The direction of the lines at the sides closer to the $q_z=0$ axis, in the limit $(C_L^v B)^2\ll \tilde{G}/K_3$, is $q_\perp/q_z \approx 1/(C_L^v B)$. These lines are well pronounced and are easily spotted already in Fig.~\ref{fig:Cvect}.
The direction of the lines at the sides closer to the $q_\perp=0$ axis, in the limit $(C_L^v B)^2\gg K_3/\tilde{G}$, is $q_\perp/q_z \approx K_3/(C_L^v B\tilde{G})$.



\subsection{Tensorial case, $C_L^t({\bf q})$}

\begin{figure}
\begin{center}
	\mbox{
	\includegraphics[width=5.5cm]{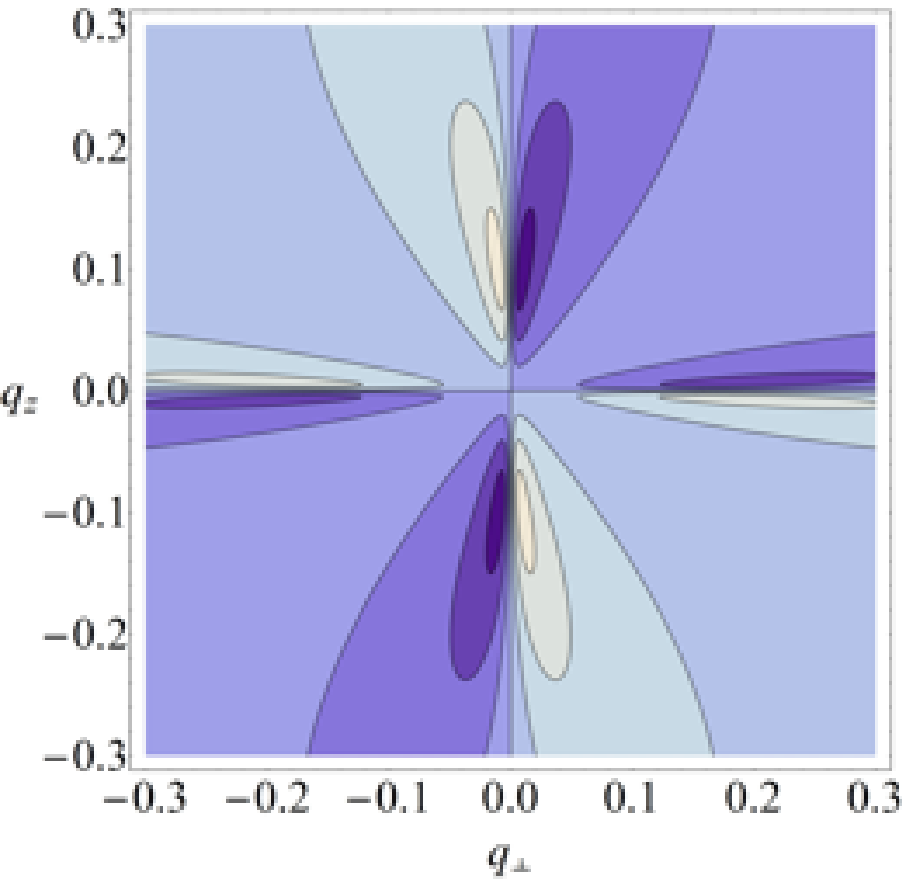}\hspace{1mm}
	\includegraphics[width=6cm]{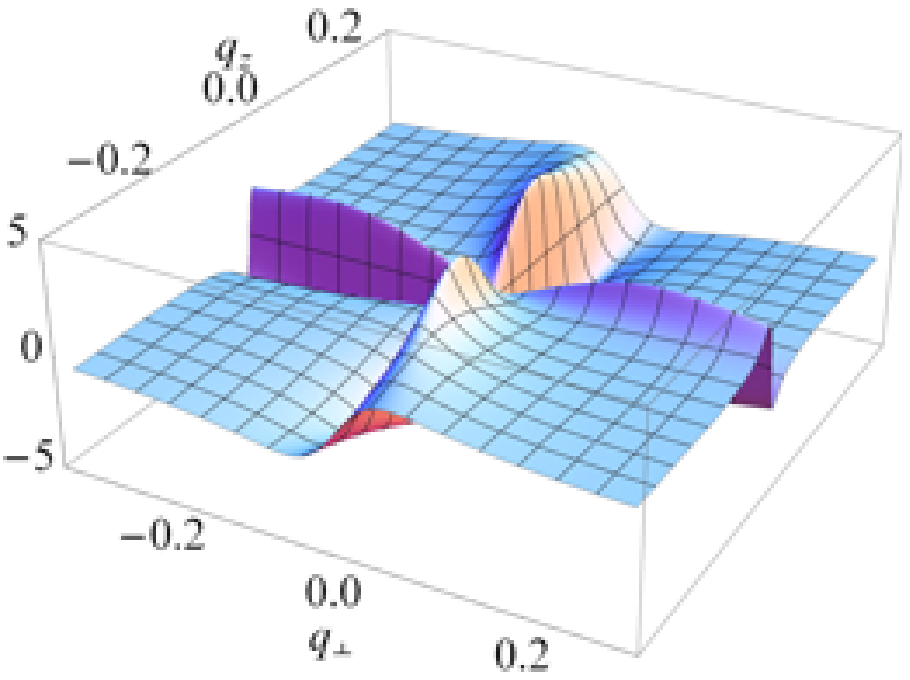}
	}
	\caption{(color online) Density -- director correlation due to the tensorial constraint $C_L^t(q_\perp, q_z)$ of Eq.~(\ref{C_Lt}) in the low-$q$ region. $K_1=0.5$, $K_3=1$, $\tilde{H}=10000$, $s_0=0.5$.}
\label{fig:Ctens}
\end{center}
\end{figure}

\noindent
In case of the tensorial constraint the connection between $C_L^t({\bf q})$ and $D_L^t({\bf q})$ is
\begin{equation}
	C_L^t({\bf q}) =-\rho_0{3s_0\, q_z q_\perp \left[(s_0+\textstyle{1\over 2})q_z^2+\textstyle{1\over 2}(1-s_0)q_\perp^2\right]
	\over {\tilde{B}/\tilde{H}}+\left[(s_0+\textstyle{1\over 2})q_z^2+\textstyle{1\over 2}(1-s_0)q_\perp^2\right]^2}\, D_L^t({\bf q})
\end{equation}
and thus with Eq.~(\ref{D_Lt}) the density -- director correlation due to the tensorial constraint is
\begin{equation}
	C_L^t({\bf q}) = -k_B T {3 s_0\, q_z q_\perp\left[(s_0+\textstyle{1\over 2})q_z^2+\textstyle{1\over 2}(1-s_0)q_\perp^2\right]\over 
	\left\{{\tilde{B}/\tilde{H}}+\left[(s_0+\textstyle{1\over 2})q_z^2+\textstyle{1\over 2}(1-s_0)q_\perp^2\right]^2\right\}\left(K_1 q_\perp^2+K_3 q_z^2\right)+\tilde{B}(3s_0\, q_z q_\perp)^2}.
	\label{C_Lt}
\end{equation}
The correlation is again zero for $H=0$. The function $C_L^t(q_\perp,q_z)$ is presented in Fig.~\ref{fig:Ctens}.
Owing to the fact that the tensorial constraint is of higher order in $q$, there exist two major differences with respect to the case of the vectorial constraint of Fig.~\ref{fig:Cvect}. $C_L^t({\bf q})$ vanishes for $q\to 0$ and has peaks at finite $q$. Furthermore, for $s_0<1$ it features secondary ridges along the $q_z=0$ direction, not existing in the vectorial case.

\begin{figure}
\begin{center}
	\includegraphics[width=6cm]{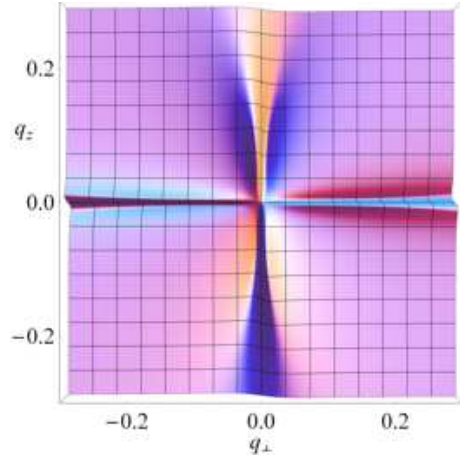}
	\caption{(color online) Top view of the tensorial density -- director correlation landscape, revealing parabolic dependence of the primary crest $q_\perp(q_z)$ and the secondary crest $q_z(q_\perp)$, respectively. $K_1=0.5$, $K_3=1$, $\tilde{H}=10000$, $s_0=0.5$.}
\label{fig:Ctens_ridges}
\end{center}
\end{figure}

In the vicinity of the origin, i.e., for $q^4\ll B/\tilde{H}$, the $C_L^t({\bf q})$ landscape is approximately hyperbolic,
\begin{equation}
	C_L^t({\bf q}\to 0) = -{k_B T\over B} {\tilde{H}\over K_3} 3s_0\, q_\perp q_z 
		{(s_0+\textstyle{1\over 2})q_z^2+\textstyle{1\over 2}(1-s_0)q_\perp^2\over q_z^2 + (K_1/K_3) q_\perp^2},
\end{equation}
where the dominant dependence is given by $q_\perp q_z$, while the last factor is merely a weak decoration not very far from unity.
It can be shown that in the region $q^4\gg B/\tilde{H}$, $K_i q^2\ll B$, the equation of the crests is 
\begin{equation}
	q_\perp \approx \pm\sqrt{K_3\over B}{s_0+{1\over 2}\over 3s_0}\, q_z^2
	\label{primary}
\end{equation}
for the primary ridges and
\begin{equation}
	q_z \approx \pm {1\over 2}\sqrt{K_1\over B}{1-s_0\over 3s_0}\, q_\perp^2
	\label{secondary}
\end{equation}
for the secondary ridges. In both cases the parabolas are well recognized when viewed from the top, Fig~\ref{fig:Ctens_ridges}.

Taking into account the connection Eq.~(\ref{primary}), the extrema of the primary ridges are at 
\begin{equation}
	q_z^4 \approx {B\over\tilde{H}}{3\over 2(s_0+\textstyle{1\over 2})^2}
\end{equation}
and their height is
\begin{equation}
	C_L^t \approx \pm{k_B T\over B}{{3^{3/4}\over 5\sqrt{2}}}{\textstyle\sqrt{s_0+{1\over 2}}}\, \left(B\tilde{H}\over K_3^2\right)^{1/4}.
\end{equation}
Taking into account the connection Eq.~(\ref{secondary}), the extrema of the secondary ridges are at 
\begin{equation}
	q_\perp^4 \approx {B\over\tilde{H}}{6\over (1-s_0)^2}
\end{equation}
and their height is
\begin{equation}
	C_L^t \approx \pm{k_B T\over B}{3^{3/4}\over 8\times 2^{1/4}}\sqrt{1-s_0}\, \left(B\tilde{H}\over K_1^2\right)^{1/4}.
\end{equation}

The ridges with extremal values of $C_L^v$ and $C_L^t$ in Figs.~\ref{fig:Cvect} and \ref{fig:Ctens} are rather sharp and narrowly spaced. Resolving them requires a good resolution in $\bf q$-space, i.e., a large system in numerical simulations. The $q$-spacing of the ridges in the region where they are prominent scales as $\sim \sqrt{B/\tilde{G}}$ for the vectorial case and as $\sim\sqrt{K_i/\tilde{H}}$ for the tensorial case. Thus, they can be more easily resolved when the constraints are weaker, but in this case their heights are lower. Preliminary investigations show that the simulations are very much in this regime.

\section{Discussion and conclusion}

\noindent
We have presented the structure factor, the director autocorrelation, and the density -- director correlation functions for a continuum worm-like description of a main-chain polymer nematic, constrained by the tensorial conservation law, all the while making comparisons with analogous correlation functions for the case of the vectorial conservation law.

In general, the correlation functions influenced by the constraints differ substantially from the correlation functions of the non-polymeric case, if the constraints are strong. In practice this requires long chains. Moreover, for the tensorial constraint to be well distinguishable from the vectorial one, the chain persistence length should be much smaller than the total length of the chain, so that hairpins (chain backfolding) are numerous and the polar order is small. 

If this is not satisfied, i.e., if chain backfolding is costly, then the tensorially-constrained configurations are expected to be similar to vectorially-constrained ones. Namely, in contrast to the first order vectorial constraint, the tensorial constraint is second order in spatial derivatives and thus possesses two families of solutions, one of which is expected to reduce to the vectorial constraint if hairpins are forbidden. 
This is indeed supported by a special case---the (rather unphysical) limit of constant density and perfect order, $s=1$, for which 
one can directly relate the splay of the director field to the local density of hairpins (c.f. Eq.~(39) of Ref.~\cite{svensek}). This indicates that in the absence of hairpins the splay is expelled, or in general, that we have recovered the vectorial conservation law with hairpins acting as sources, thus being physically equivalent to chain heads and tails. Hence in this and only this limit the tensorial constraint is reduced to the vectorial one.
General aspects of this connection are however not yet thoroughly understood and further investigations are being carried out to gain more insight into the various ramifications of the vectorial and tensorial conservation laws and their interpretation.

It furthermore follows from our mesoscopic free energy functional, Eq.~(\ref{ft_q}), that microscopic chain backfolding is not penalized and therefore contributes to minimize the energy cost of the constrained density and director variations. As a consequence, in our model the tensorially-constrained configurations with chain backfolding are therefore always different from the vectorially-constrained configurations, which exhibit no chain backfolding by construction. In other words, within our model based on the functionals Eqs.~(\ref{Fnem})-(\ref{F_t}), we are always in the flexible chain limit. 

For rather short chains, i.e., values of $\tilde{G}q^2/B\sim 1$ and $\tilde{H}q^4/B\sim 1$, the effects of both constraints are less pronounced and consequently it is not easy to distinguish between the vectorial and tensorial constraints within this regime, as can be clearly observed in Figs.~\ref{fig:Svect}, \ref{fig:Stens}, \ref{fig:Cvect}, and \ref{fig:Ctens} in the vicinity of the origin. Qualitatively, the vectorial and tensorial structure factors are very similar in that region of the parameter space  and can be distinguished only by a detailed analysis of the functional dependence of $1/S$ as indicated in Secs.~\ref{sec:parallel} and \ref{sec:perpendicular}.
Furthermore, in this parameter regime the director autocorrelation of the tensorial case $D_L^t$ can be hardly distinguished from the non-polymeric $D_L$. The difference seems to be more discernible in the vectorial case: the peak of $D_L^v$ is still elongated along the $q_z$ axis, whereas the peak of non-polymeric $D_L$ is elongated along the $q_\perp$ axis in the case of the more common elastic anisotropy $K_3>K_1$.

In principle, the most distinct feature setting apart the two types of constraints, that persists also in the regime of weak constraints, is the behavior of the density -- director correlation functions at ${\bf q}=0$: $C_L^v(0)$ is finite, whereas $C_L^t(0)$ vanishes. The difference here is thus qualitative. But in order that it would become actually manifest, we must compare a system with polar orientational order without hairpins on one side, with a system with quadrupolar orientational order with abundant hairpins. The latter case can be rather difficult to achieve, especially in simulations, as it requires large systems.
With limited chain backfolding, however, one presumably cannot observe more than smaller or larger deviations from $C_L^v$, depending on the abundance of hairpins. In this case, near ${\bf q}=0$ the deviations from $C_L^v$, which is finite, towards $C_L^t$, which is vanishing, could be rather delicate to trace.

\begin{acknowledgments}
\noindent
The Authors acknowledge the support of the Slovenian Research Agency (Grants No. J1-4297, J1-4134, N1-0019, P1-0055, P1-0099). Many thanks to K. Ch. Daoulas, G. Ska\v cej, A. Popadi\' c and M. Praprotnik for stimulating discussions and interest in the subject.
\end{acknowledgments}

\end{document}